\documentclass{ws-procs9x6}
\usepackage{braket}
\begin{document}

\title{
QUANTUM ANNEALING AND QUANTUM FLUCTUATION EFFECT IN FRUSTRATED ISING SYSTEMS
}

\author{SHU TANAKA}

\address{
Department of Chemistry, University of Tokyo,\\
7-3-1, Hongo, Bunkyo-ku, Tokyo, 113-0033, Japan\\
E-mail: shu-t@chem.s.u-tokyo.ac.jp
}

\author{RYO TAMURA}

\address{
Institute for Solid State Physics, University of Tokyo,\\
5-1-5, Kashiwanoha, Kashiwa-shi, Chiba, 277-8501, Japan\\
E-mail: r.tamura@issp.u-tokyo.ac.jp
}

\begin{abstract}
Quantum annealing method has been widely attracted attention in statistical physics and information science since it is expected to be a powerful method to obtain the best solution of optimization problem as well as simulated annealing.
The quantum annealing method was incubated in quantum statistical physics.
This is an alternative method of the simulated annealing which is well-adopted for many optimization problems.
In the simulated annealing, we obtain a solution of optimization problem by decreasing temperature (thermal fluctuation) gradually.
In the quantum annealing, in contrast, we decrease quantum field (quantum fluctuation) gradually and obtain a solution.
In this paper we review how to implement quantum annealing and show some quantum fluctuation effects in frustrated Ising spin systems.

\end{abstract}


\bodymatter

\section{Introduction}

In information science, it has been an important issue to propose a novel method or improve known methods for obtaining the best solution of optimization problems.
Optimization problem is to find the state in which the real-valued cost/gain function has the minimum/maximum value.
In general, since there are huge number of elements in optimization problems, we cannot obtain the best solution by a naive method such as full search in limited time and present resources.
This is because the number of states increases exponentially with the number of elements.
Optimization problems are often represented by binary language of $0$s and $1$s.
Here the minimum unit of information is called ``bit''.
This situation is the same as the Ising model\cite{STIsing-1925}.
In the Ising model, the minimum unit corresponds to ``spin'' which can have the value $+1$ or $-1$.
Then, in many cases, optimization problems can be represented by the Ising model with random interaction $J_{ij}$.
The Hamiltonian is represented as 
\begin{eqnarray}
 \label{STeq:Ham_Ising}
 {\cal H}_{\rm Ising} = - \sum_{i,j} J_{ij} \sigma_i^z \sigma_j^z,
  \quad
  \sigma_i^z = \pm 1.
\end{eqnarray}
Here $\sigma_i^z$ is an assigned integer $\pm 1$ 
(not operator\footnote{
Here we do not consider quantum system.
In quantum systems $\hat{\sigma}_i^z$ represents the $z$-component of the Pauli matrix of the $i$-th site.
Even if we consider the quantum version of the Hamiltonian given by Eq.~(\ref{STeq:Ham_Ising}), there is no off-diagonal elements.
In this paper we discriminate $c$-number and matrix (operator) without or with the hat $\hat{}$.
}) 
at the $i$-th site.
If an optimization problem is mapped onto the Ising Hamiltonian given by Eq.~(\ref{STeq:Ham_Ising}), the cost function of the optimization problem is expressed as the internal energy of the mapped Ising spin system.
Thus, to find the best solution of optimization problem corresponds to obtain the ground state in the mapped Ising spin system.

Ising models with random interactions often have a difficulty of slow relaxation at low temperature. 
In general, energy landscape of complicated Ising model prevents the system from relaxation\cite{STMezard-1987,STFischer-1993,STYoung-1998}.
In order to avoid such a difficulty, many methods have been proposed in statistical physics.
The most famous one is simulated annealing which was proposed by Kirkpatrick {\it et al}\cite{STKirkpatrick-1983,STKirkpatrick-1984}.
In the simulated annealing, we decrease temperature (thermal fluctuation) gradually and obtain a solution of optimization problem.
In order to find the best solution, we should sample states according to transition probability defined in some way.
Transition probability from a state $\Sigma$ to the other state $\Sigma'$ is often defined so as to be proportional to ${\rm e}^{\beta(E(\Sigma)-E(\Sigma'))}$, where $E(\Sigma)$ is the eigenenergy of the eigenstate $\Sigma$ and $\beta$ is an inverse temperature $\beta = T^{-1}$.
Here we set the Boltzmann constant $k_{\rm B}$ to be unity.
At low temperature, because of large $\beta$, a transition between states does not occur with frequency.
On the other hand, at high temperature, the probability distribution is almost flat.
Then a transition from a certain state to the other state often occurs.
By using the effect of thermal fluctuation, we are easy to obtain the best solution by decreasing temperature gradually.
Actually, Kirkpatrick {\it et al.} demonstrated finding a stable state of the random Ising spin systems.
After that, the simulated annealing is widely adopted for many kinds of optimization problems in physics, chemistry, and information science since the simulated annealing is easy to implement.
Moreover it has been proved that the simulated annealing can find the best solution definitely when we decrease temperature slow enough\cite{STGeman-1984}.
Then the simulated annealing has been regarded as guaranteed general method for optimization problems.

Since there are many types of optimization problems, there have been proposed corresponding methods in order to obtain the best solution of individual optimization problem.
This is a style of information engineering.
However it is also important to construct a general method for optimization problems such as the simulated annealing.
Based on this point of view, the quantum annealing has been also developed\cite{STFinnila-1994,STKadowaki-1998,STBrooke-1999,STFarhi-2001,STSantoro-2002,STMatronak-2004,STBattaglia-2005,STSuzuki-2005,STDas-2005,STTanaka-2007,STDas-2008,STKurihara-2009,STSato-2009,STMorita-2009,STTanaka-2009,STInoue-2010,STTanaka-2010,STTanaka-2010book,STChandra-2010,STOhzeki-2011,STTanaka-2011a,STTanaka-2011b,STTanaka-2011c}.
In the quantum annealing, instead of decreasing temperature (thermal fluctuation), we decrease quantum field (quantum fluctuation) gradually and obtain a solution of optimization problem.
The quantum annealing method has been expected as a powerful tool to obtain the best solution of optimization problem.
However there are few examples that the efficiency of quantum annealing is worse than that of the simulated annealing.
In order to know when to use the quantum annealing, it is a significant issue to investigate microscopic nature of quantum field response in simple models.

The organization of the rest of the paper is as follows.
In Section 2, we review implementation methods of quantum annealing.
In Section 3, we consider quantum field response of frustrated Ising spin systems.
In Section 4, we conclude this paper briefly.

\section{Implementation Methods of Quantum Annealing}

In this section, we review concept of quantum annealing and implementation methods.
The quantum annealing is expected to be an alternative method of simulated annealing and a general framework for optimization problems as mentioned above.
Here we assume that our target optimization problem can be expressed by the Ising model given as
\begin{eqnarray}
 \label{STeq:Hamclassical}
 {\cal H}_{\rm c} = - \sum_{i,j} J_{ij} \hat{\sigma}_i^z \hat{\sigma}_j^z,
\end{eqnarray}
where $\hat{\sigma}_i^\alpha$ denotes the $\alpha$-component of the Pauli matrix defined by
\begin{eqnarray}
 \hat{\sigma}_i^x = 
  \left(
   \begin{array}{cc}
    0 & 1 \\
    1 & 0
   \end{array}
  \right),
  \quad
  \hat{\sigma}_i^y = 
  \left(
   \begin{array}{cc}
    0 & -i \\
    i & 0
   \end{array}
  \right),
  \quad
  \hat{\sigma}_i^z = 
  \left(
   \begin{array}{cc}
    1 & 0 \\
    0 & -1
   \end{array}
  \right).
\end{eqnarray}
Equation (\ref{STeq:Hamclassical}) is a quantum version of the Hamiltonian given by Eq.~(\ref{STeq:Ham_Ising}) and a diagonal matrix.
Our purpose is to find the ground state of the Hamiltonian given by Eq.~(\ref{STeq:Hamclassical}).
In order to perform quantum annealing, we introduce an additional time-dependent Hamiltonian ${\cal H}_{\rm q}(t)$ which represents quantum fluctuation.
Then the total Hamiltonian is expressed as 
\begin{eqnarray}
 \label{STeq:generalQA}
 {\cal H}(t) = {\cal H}_{\rm c} + {\cal H}_{\rm q}(t).
\end{eqnarray}
Hereafter we refer to ${\cal H}_{\rm c}$ and ${\cal H}_{\rm q}(t)$ as classical Hamiltonian and quantum Hamiltonian, respectively.
The most typical form of the quantum Hamiltonian for the Ising spin system is transverse field given as
\begin{eqnarray}
 \label{STeq:TofI}
 {\cal H}_{\rm q}(t) = -\Gamma(t) \sum_i \hat{\sigma}_i^x.
\end{eqnarray}
Although there is an ambiguity of ${\cal H}_{\rm q}(t)$, we concentrate on the case that ${\cal H}_{\rm q}(t)$ is the form given by Eq.~(\ref{STeq:TofI}) in this paper.

\begin{figure}[b]
 \begin{center}
  \psfig{file=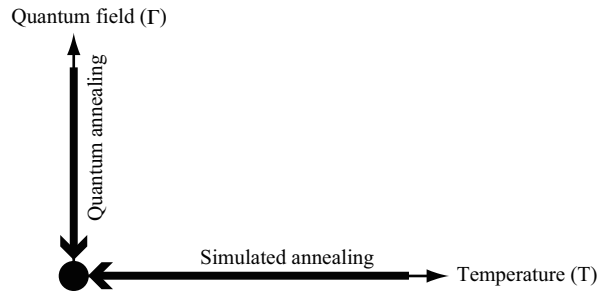,scale=1.0}
 \end{center}
 \caption{
 Schematic diagram of simulated annealing and quantum annealing.
 The aim of annealing methods is to obtain the best solution at the circle ($T=\Gamma=0$).
 In the simulated annealing, we decrease temperature $T$ gradually from high temperature to (nearly) zero temperature.
 In the quantum annealing, we decrease quantum field $\Gamma$ gradually from large value to (nearly) zero.
 }
 \label{STfig:schematicSAQA}
\end{figure}

In the quantum annealing, we gradually decrease quantum field $\Gamma$ and obtain the final state at $\Gamma=0$, whereas we gradually decrease temperature $T$ and obtain the final state at $T=0$ in the simulated annealing (Fig.~\ref{STfig:schematicSAQA}).
Recently hybrid-type quantum simulated annealing in which we decrease temperature and quantum field simultaneously has developed\cite{STKurihara-2009,STSato-2009}.
Quantum annealing can be categorized by three types summarized in Table \ref{STtable:methods}.
There are some implementation methods of quantum annealing theoretically and experimentally.
We can divide into two types of implementation theoretical methods according to how to treat time-development.
The first one is a stochastic method which is realized by the Monte Carlo method.
The Monte Carlo methods have been widely adopted for analysis of equilibrium state of many-body strongly correlated systems in statistical physics.
Since in some cases we face on the difficulty of obtaining the equilibrium state by naive Monte Carlo methods, many efficient algorithms have been developed\cite{STSwendsen-1987,STWolff-1989,STHukushima-1996,STHarada-2004,STNakamura-2008,STMorita-2009,STSuwa-2010,STSuwa-2011}.
Then if we adopt these engineered methods depending on the situation, we can use the Monte Carlo method as a powerful tool for quantum annealing.

The other type of theoretical methods is a deterministic methods.
We can calculate real-time dynamics by solving the time-dependent Schr\"odinger equation directly\cite{STKadowaki-1998,STBrooke-1999,STFarhi-2001,STSantoro-2002,STTanaka-2007,STTanaka-2009} or performing the time-dependent density matrix renormalization group\cite{STSuzuki-2007}.
There is another type of deterministic method, which is mean-field type calculation\cite{STTanaka-2000}.
Moreover, variational Bayes inference which is a useful method in information science can be regarded as a method based on mean-field calculation.
This method can also treat large-scale problems as well as the Monte Carlo method.
Originally, the variational Bayes inference has been adopted for many optimization problems.
Recently a quantum annealing version of variational Bayes inference was developed\cite{STSato-2009}.

Quantum annealing can realize not only theoretically in classical computer but also experimentally.
Artificial lattices such as optical lattice have been made in order to simulate strongly correlated fermionic systems and spin systems\cite{STBloch-2008,STKim-2010,STMa-2011,STSimon-2011,STStruck-2011}.
Since in artificial lattice systems, we can control parameters in the Hamiltonian, it is expected that these systems become a ``quantum computer'' which can find the ground state of complicated systems.
\begin{table}[t]
\tbl{
 Quantum annealing is categorized by three types depending on how to treat time-development.
}
{
\begin{tabular}{@{}ccc@{}}\toprule
 \multicolumn{2}{c}{Theoretical methods}& Experiments\\
 Stochastic methods & Deterministic methods &  \\
 \colrule
 Quantum Monte Carlo & Real-time dynamics & Artificial lattices \\
 & (Exact diagonalization,& (Optical lattice)\\
 & time-dependent DMRG) & \\
 & Mean-field type method & \\
 \botrule
\end{tabular}
}
 \label{STtable:methods}
\end{table}

In this paper we focus on two theoretical implementation methods.
In Sec.~\ref{STsec:qmc}, we review the quantum annealing method for the transverse Ising model by using the quantum Monte Carlo method.
In Sec.~\ref{STsec:realtime}, we explain the real-time dynamics using the Schr\"odinger equation for the transverse Ising model.

\subsection{Quantum Monte Carlo method}
\label{STsec:qmc}

In this subsection, we explain how to implement quantum annealing by the quantum Monte Carlo method.
Here we take the transverse Ising model as an example.
The quantum Monte Carlo method is a useful tool to obtain thermodynamic properties of strongly correlated many-body systems such as quantum spin systems, bosonic systems, and fermionic systems.
Before we review the quantum Monte Carlo method for the transverse Ising model, we show implementation method of classical Monte Carlo method for Ising spin systems without transverse field given by Eq.~(\ref{STeq:Ham_Ising}).
The main procedure of the classical Monte Carlo method is as follows.
\begin{description}
 \item[Step 1] We prepare an initial state\footnote{There is an ambiguity how to prepare the initial state.
	    For example, we can use both random spin configuration and completely polarized configuration.}.
 \item[Step 2] We select a spin randomly.
 \item[Step 3] We flip the selected spin with probability defined by some way.
 \item[Step 4] We repeat Step 2 and Step 3 until physical quantities converge.
\end{description}
Performance of Monte Carlo method depends on how to choice the transition probability in Step 3.
The simplest definition of the transition probability is the Metropolis method.
Let $\Sigma$ and $\Sigma'$ be the state before and after spin flip, respectively.
In the Metropolis method, we change the state from $\Sigma$ to $\Sigma'$ according to the following probability $P(\Sigma'|\Sigma)$:
\begin{eqnarray}
 P(\Sigma'|\Sigma) = 
  \begin{cases}
   1 & \qquad (E(\Sigma') \le E(\Sigma))\\
   {\rm e}^{\beta(E(\Sigma)-E(\Sigma'))} & \qquad (E(\Sigma') > E(\Sigma))
  \end{cases},
\end{eqnarray}
where $E(\Sigma)$ is the eigenenergy of the eigenstate $\Sigma$.
In this rule, when the eigenenergy of the candidate state $\Sigma'$ is lower than that of the present state, the state changes from $\ket{\Sigma}$ to $\ket{\Sigma'}$ definitely.
Here $\ket{\Sigma}$ represents the direct product of $N$-spin states as follows:
\begin{eqnarray}
 \ket{\Sigma} = \ket{\sigma_1} \otimes \ket{\sigma_2} \otimes \cdots \otimes \ket{\sigma_N},
\end{eqnarray}
where $\ket{\sigma_i}$ denotes the state at the $i$-th site.
It should be noted that we can easily calculate $E(\Sigma)$ since now we consider the Hamiltonian in which there is no off-diagonal element.

Next we review the quantum Monte Carlo method for the transverse Ising model given by
\begin{eqnarray}
 {\cal H} = -\sum_{i,j} J_{ij} \hat{\sigma}_i^z \hat{\sigma}_j^z - \Gamma \sum_i \hat{\sigma}_i^x.
\end{eqnarray}
In this model for large number of spins, it is difficult to obtain all eigenenergies and eigenstates in practice because of off-diagonal elements in the Hamiltonian.
Then we cannot use the Monte Carlo method directly since we should know all eigenvalues and corresponding eigenstates.
We have to consider an alternative representation which enables us to treat this model.
In general, partition function of $d$-dimensional transverse Ising model (quantum system) is equivalent to that of $(d+1)$-dimensional Ising model without transverse field (classical system).
This correspondence can be derived by the path-integral representation (in other words, the Trotter-Suzuki decomposition\cite{STTrotter-1959,STSuzuki-1976}).

\begin{figure}[t]
 \begin{center}
  \psfig{file=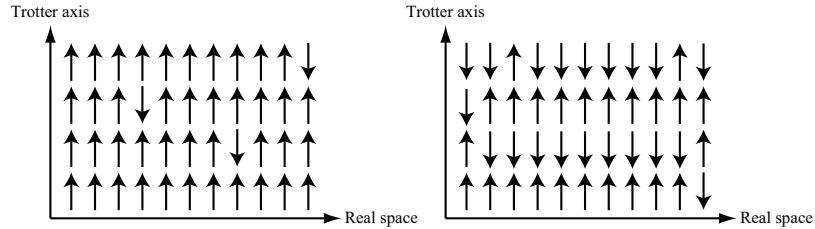,scale=0.8}
 \end{center}
 \caption{
 Schematic spin configurations of the transverse Ising chain given by Eq.(\ref{STeq:transverseIsingchain}) for small $\Gamma$ (left panel) and large $\Gamma$ (right panel).
 }
 \label{STfig:snapshots}
\end{figure}

Then we can use the Monte Carlo method for the transverse Ising model by considering path-integral representation.
For simplicity, in order to demonstrate path-integral representation, we consider the ferromagnetic Ising chain with transverse field whose Hamiltonian is given by
\begin{eqnarray}
 \label{STeq:transverseIsingchain}
 {\cal H} = {\cal H}_{\rm c} + {\cal H}_{\rm q},
  \quad
  {\cal H}_{\rm c} = -J \sum_{i=1}^{N} \hat{\sigma}_i^z \hat{\sigma}_{i+1}^z,
  \quad
  {\cal H}_{\rm q} = -\Gamma \sum_{i=1}^N \hat{\sigma}_i^x,
\end{eqnarray}
where the periodic boundary is assumed: $\sigma_{N+1}=\sigma_1$.
At the temperature $T(=\beta^{-1})$, the partition function of this system is expressed as 
\begin{eqnarray}
 Z = {\rm Tr}\, {\rm e}^{-\beta {\cal H}} = {\rm Tr}\, {\rm e}^{-\beta({\cal H}_{\rm c}+{\cal H}_{\rm q})}
= \sum_\Sigma \Braket{\Sigma|{\rm e}^{-\beta({\cal H}_{\rm c}+{\cal H}_{\rm q})}|\Sigma}.
\end{eqnarray}
We cannot obtain the partition function by this direct expression since $\braket{\Sigma|{\rm e}^{-\beta({\cal H}_{\rm c}+{\cal H}_{\rm q})}|\Sigma}$ is not tractable.
Then by using integer $m$, we decompose the matrix exponential ${\rm e}^{-\beta({\cal H}_{\rm c}+{\cal H}_{\rm q})}$ as follows.
\begin{eqnarray}
 \exp\left[
      -\beta({\cal H}_{\rm c}+{\cal H}_{\rm q})
     \right]
 = \exp\left[
	{\rm e}^{-\frac{1}{m}\beta{\cal H}_{\rm c}}
	{\rm e}^{-\frac{1}{m}\beta{\cal H}_{\rm q}}
       \right]^m
 + {\cal O}\left( \left( \frac{\beta}{m}\right)^2\right).
\end{eqnarray}
Then we obtain the partition function by using $m$:
\begin{eqnarray}
 \nonumber
 Z &&= \sum_\Sigma \Braket{\Sigma|{\rm e}^{-\beta({\cal H}_{\rm c}+{\cal H}_{\rm q})}|\Sigma}\\
\nonumber   
&&= \sum_{\Sigma_1,\Sigma_1',\cdots,\Sigma_m,\Sigma_m'}
    \Braket{\Sigma_1|{\rm e}^{-\frac{\beta}{m}{\cal H}_{\rm c}}|\Sigma_1'}
    \Braket{\Sigma_1'|{\rm e}^{-\frac{\beta}{m}{\cal H}_{\rm q}}|\Sigma_2}\\
\nonumber  
&&\times 
    \Braket{\Sigma_2|{\rm e}^{-\frac{\beta}{m}{\cal H}_{\rm c}}|\Sigma_2'}
    \Braket{\Sigma_2'|{\rm e}^{-\frac{\beta}{m}{\cal H}_{\rm q}}|\Sigma_3}\\
\nonumber 
&&\times \cdots \\
 && \times
    \Braket{\Sigma_m|{\rm e}^{-\frac{\beta}{m}{\cal H}_{\rm c}}|\Sigma_m'}
    \Braket{\Sigma_m'|{\rm e}^{-\frac{\beta}{m}{\cal H}_{\rm q}}|\Sigma_1},
\end{eqnarray}
where $\ket{\Sigma_k}$ represents the direct product of $N$ spin states as well as $\ket{\Sigma}$:
\begin{eqnarray}
 \ket{\Sigma_k} = \ket{\sigma_{k,1}} \otimes \ket{\sigma_{k,2}} \otimes \cdots \otimes \ket{\sigma_{k,N}}.
\end{eqnarray}
The index $k$ represents the coordinate along the Trotter axis.
Hereafter we refer to $m$ as the Trotter number.
Since the classical Hamiltonian ${\cal H}_{\rm c}$ is a diagonal matrix, the following relation is satisfied:
\begin{eqnarray}
 \hat{\sigma}_j^z \ket{\Sigma_k} = \sigma_{k,j}^z \ket{\Sigma_k}.
\end{eqnarray}
From the above relation and simple calculation,
\begin{eqnarray}
 \nonumber 
  &&\Braket{\Sigma_k | {\rm e}^{-\frac{\beta{\cal H}_{\rm c}}{m}}|\Sigma_k'}
  = \exp{\left( \frac{\beta J}{m} \sum_{i=1}^{N} \sigma_{k,i}^z \sigma_{k,i+1}^z \right)}
  \prod_{i=1}^N \delta(\sigma_{k,i}^z,\sigma_{k,i}^{z'}),
  \\
  \nonumber
 &&\Braket{\Sigma_k' | {\rm e}^{-\frac{\beta{\cal H}_{\rm q}}{m}}|\Sigma_{k+1}}
  = \left[
     \frac{1}{2} \sinh \left( \frac{2\beta\Gamma}{m} \right)
    \right]^{\frac{N}{2}}
  \exp
  \left[
   \frac{1}{2} \log \coth 
   \left(
    \frac{\beta\Gamma}{m} \sum_{i=1}^N \sigma_{k,i}^{z'} \sigma_{k+1,i}^z
   \right)
  \right],
\end{eqnarray}
are obtained.
Then the partition function is expressed by the following way:
\begin{eqnarray}
 \nonumber
 Z &=& \lim_{m \to \infty}
  \left[
   \frac{1}{2} \sinh \left( \frac{2\beta\Gamma}{m}\right)
  \right]^{\frac{N}{2}}\\
 \nonumber
  &\times& \sum_{\{ \sigma_{k,i} = \pm 1\}}
   \exp 
   \left[
    \sum_{i=1}^N \sum_{k=1}^m
   \left( 
    \frac{\beta J}{m} \sigma_{k,i}^z \sigma_{k,i+1}^z 
    + \frac{1}{2} \log \coth \left( \frac{\beta\Gamma}{m} \right)
    \sigma_{k,i}^z \sigma_{k+1,i}^z
   \right)
  \right]\\
 &=& \lim_{m \to \infty} A
  \sum_{\{ \sigma_{k,i} = \pm 1\}}
  \exp (-\beta {\cal H}_{\rm eff}),
\end{eqnarray}
where $A$ is just a coefficient which is irrelevant for thermodynamic properties and the effective Hamiltonian ${\cal H}_{\rm eff}$ is defined as
\begin{eqnarray}
 {\cal H}_{\rm eff} = \sum_{i=1}^{N} \sum_{k=1}^m 
  \left[
  -\frac{J}{m} \sigma_{k,i}^z \sigma_{k,i+1}^z
  - \frac{1}{2\beta}\log \coth \left( \frac{\beta\Gamma}{m} \right)
  \sigma_{k,i}^z \sigma_{k+1,i}^z
  \right].
\end{eqnarray}
This relation means the partition function of one-dimensional transverse Ising model is equivalent to that of two-dimensional Ising model without transverse field.
Since physical quantities can be calculated by the partition function in general, we can obtain thermodynamic properties of transverse Ising chain by considering that of two-dimensional classical Ising model.
It should be noted that the abovementioned derivation does not depend on spatial dimension.
Then the method can be adopted for general Ising model with transverse field.
Figure \ref{STfig:snapshots} displays schematic spin states of the transverse Ising chain for small $\Gamma$ and large $\Gamma$ at low temperature.
When the transverse field $\Gamma$ is small, spins are almost aligned along both real space and the Trotter axis.
On the other hand, spins are aligned along only real space when the transverse field $\Gamma$ is large.
This is a nature which comes from quantum fluctuation effect.

So far, we review quantum Monte Carlo method as a tool to obtain the thermodynamic equilibrium properties.
This method can be also regarded as a realization method of stochastic dynamics.
Then we can use quantum Monte Carlo method as a method to perform quantum annealing.
In the quantum annealing, the transverse field decreases against Monte Carlo step.
The efficiency of quantum annealing by quantum Monte Carlo method has been studied, and the quantum annealing method has been succeeded to obtain not so bad solution of some optimization problems.

\subsection{Real-Time Dynamics}
\label{STsec:realtime}

\begin{figure}[b]
 \begin{center}
  \psfig{file=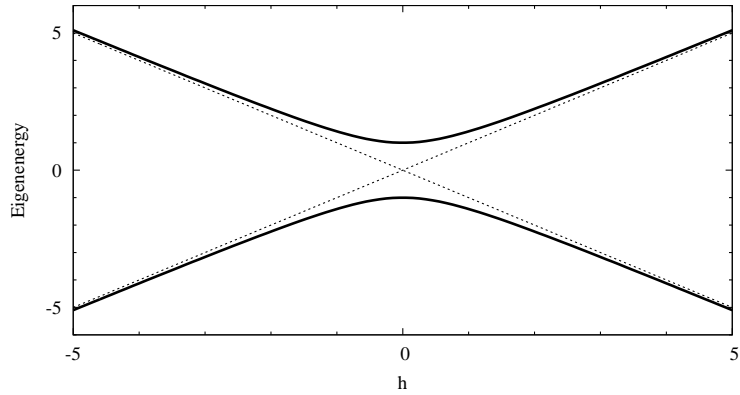,scale=0.8}
 \end{center}
 \caption{
 Eigenenergies of the single spin with longitudinal and transverse magnetic fields.
 The dotted lines and the solid curves indicate the eigenenergies for $\Gamma=0$ and $\Gamma=1$, respectively.
 }
 \label{STfig:energydiagram_single}
\end{figure}

In the previous subsection, we explained implementation method of quantum annealing by quantum Monte Carlo method.
In the method, time-evolution is treated as a stochastic process.
In this subsection, we review a method in which time-evolution is considered as a deterministic process.
There are a couple of methods which realize deterministic time-evolution.
In this paper, we focus on real-time dynamics which is derived from the Schr\"odinger equation:
\begin{eqnarray}
 i\hbar \frac{\partial}{\partial t} \ket{\psi(t)} = {\cal H}(t) \ket{\psi(t)},
\end{eqnarray}
where $\ket{\psi(t)}$ and ${\cal H}(t)$ denote time-dependent state and Hamiltonian, respectively, at time $t$.
In this method, once we prepare a state as the initial state, the time-development of the state is determined.
Then, how to choice the initial state is important.
In the sense of quantum annealing, it is a usual way to prepare the initial state which can be made easily.
In the case of the transverse Ising model, fully polarized state is a trivial state at large transverse field.
The state is expressed as 
\begin{eqnarray}
 \ket{\psi(0)} = \ket{\rightarrow \cdots \rightarrow},
\end{eqnarray}
where the state $\ket{\rightarrow}$ is defined as
\begin{eqnarray}
 \ket{\rightarrow} = \frac{1}{\sqrt{2}} \left( \ket{\uparrow} + \ket{\downarrow}\right).
\end{eqnarray}
The state $\ket{\rightarrow}$ is an eigenstate of the $x$-component of the Pauli matrix $\hat{\sigma}^x$.
If the energy level of the prepared initial state does not cross with other levels for all transverse field strength $\Gamma$, we can obtain the ground state in the adiabatic limit.
However, in practice, since we decrease the transverse field with finite speed, transition to excited levels is inevitable.
In order to consider such a nonadiabatic transition, we study a single spin system with longitudinal and transverse fields.
The Hamiltonian is given by
\begin{eqnarray}
 {\cal H}_{\rm single} = -h \hat{\sigma}^z -\Gamma \hat{\sigma}^x = 
  \left(
   \begin{array}{cc}
    -h & -\Gamma\\
    -\Gamma & h\\
   \end{array}
  \right).
\end{eqnarray}
Eigenvalues of this Hamiltonian are
\begin{eqnarray}
 E_{\pm} = \pm \sqrt{h^2 + \Gamma^2}.
\end{eqnarray}
Figure \ref{STfig:energydiagram_single} denotes the eigenenergies as functions of longitudinal field $h$.
Then the energy difference (energy gap) between two eigenstates is $2\sqrt{h^2+\Gamma^2}$.
The energy gap takes the minimum value $2\Gamma$ at $h=0$ for $\Gamma$.

Next we consider a single spin problem with time-dependent longitudinal field $h(t)=vt$ and the fixed transverse field $\Gamma$:
\begin{eqnarray}
 {\cal H}_{\rm single}(t) = - vt \hat{\sigma}^z - \Gamma \hat{\sigma}^x = 
  \left(
   \begin{array}{cc}
    -vt & -\Gamma \\
    -\Gamma & vt
   \end{array}
  \right).
\end{eqnarray}
Here we set the down state $\ket{\downarrow}$ as the initial state.
This state is the ground state of the Hamiltonian in the limit of $t\to -\infty$.
When the transverse field $\Gamma$ is zero, the state $\ket{\downarrow}$ is the eigenstate of the Hamiltonian.
From the viewpoint of quantum annealing, if the symmetry of the prepared initial state is different from that of the ground state at the final time, we cannot obtain the ground state at all.
We next consider the case for finite transverse field $\Gamma$.
In this case, the energy level of the ground state does not cross with that of the excited state as stated before.
Then the state becomes $\ket{\uparrow}$ at $t\to\infty$ in the adiabatic limit ($v\to 0$).
When we sweep longitudinal magnetic field with finite speed $v$, linear combination of eigenstates is obtained even at $t\to\infty$.

\begin{figure}[b]
 \begin{center}
  \psfig{file=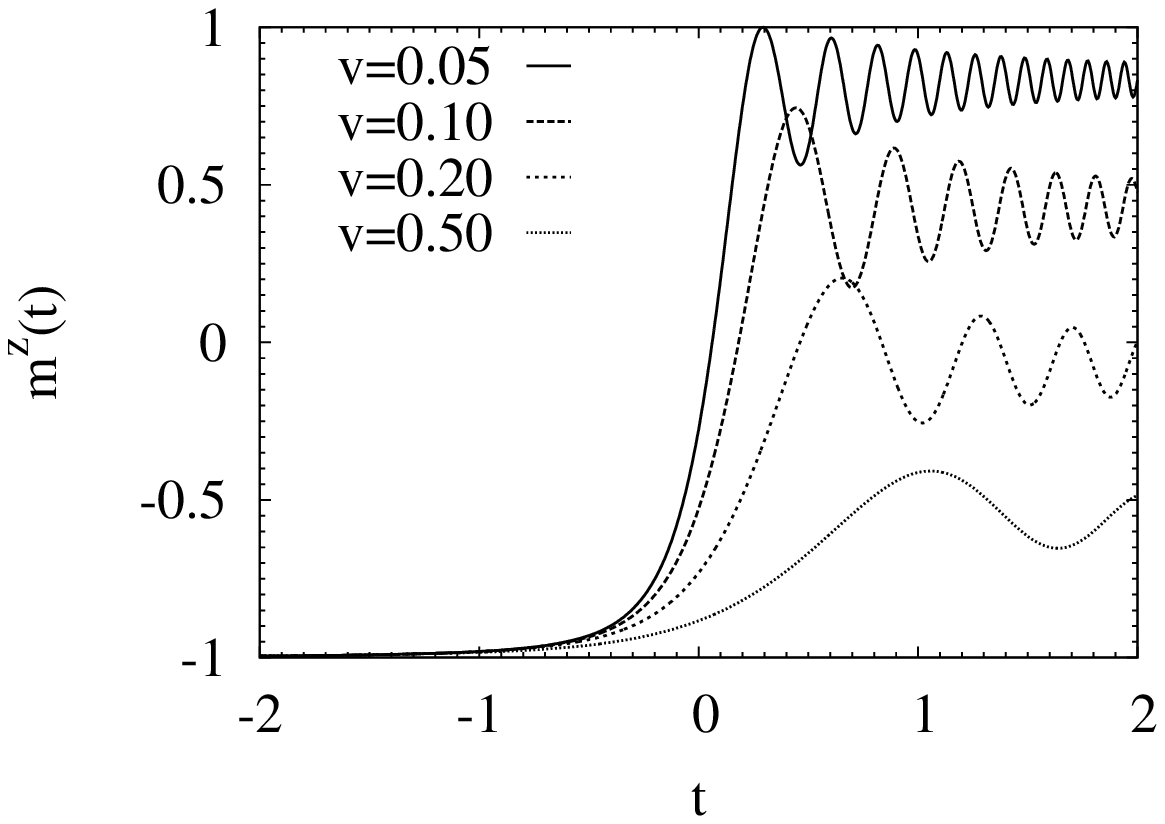,scale=0.4}
  \psfig{file=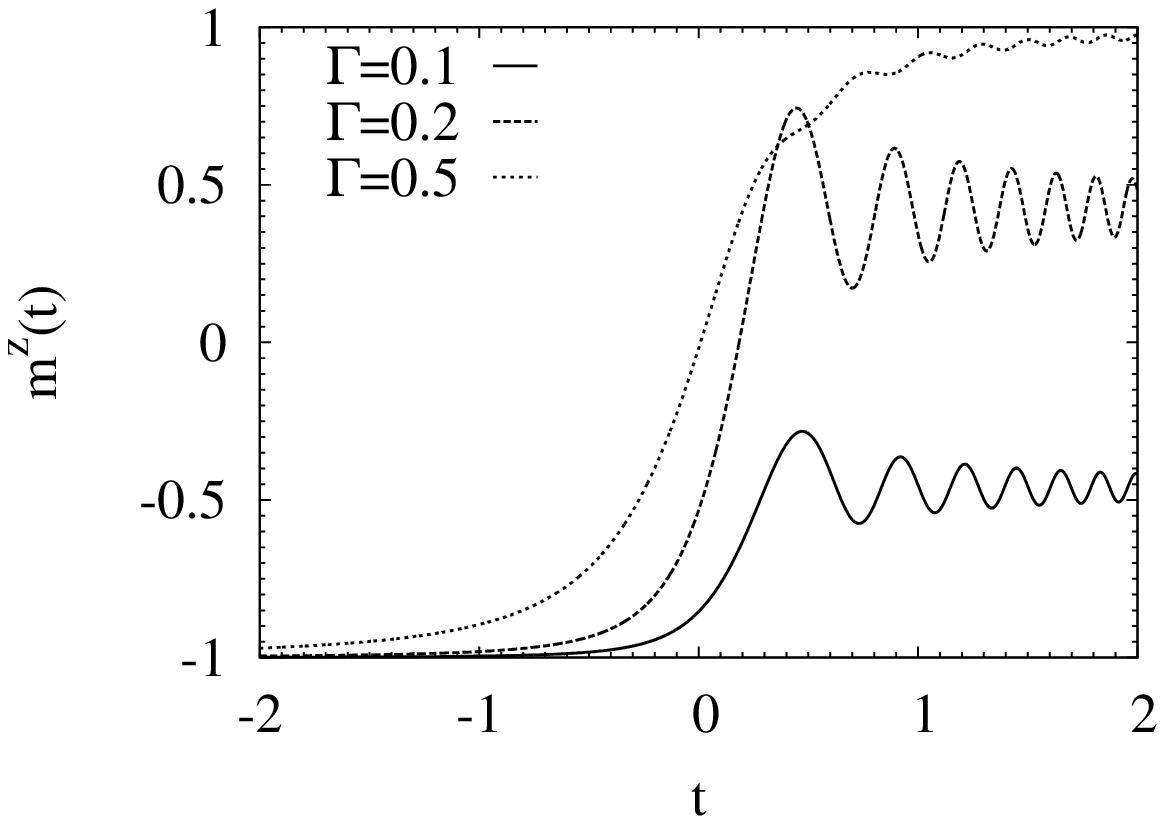,scale=0.4}
 \end{center}
 \caption{
 Time-development of the magnetization along the $z$-axis for various sweeping speed under fixed $\Gamma=0.2$ (left panel) and for various ${\Gamma}$ under fixed sweeping speed $v=0.1$ (right panel).
 }
 \label{STfig:LZS}
\end{figure}

The left and right panels in Fig.~\ref{STfig:LZS} show time-development of magnetization along the $z$-axis for various sweeping speed with fixed transverse field $\Gamma=0.2$ and for various $\Gamma$ with fixed sweeping speed $v=0.1$, respectively.
Here the magnetization along the $z$-axis is calculated by
\begin{eqnarray}
 m^z(t) = \braket{\psi(t)|\hat{\sigma}^z|\psi(t)}.
\end{eqnarray}
As the sweeping speed becomes slow and/or the energy gap becomes large, the state approaches the adiabatic limit of the ground state ($m^z(t=+\infty)=+1$).
This is called the Landau-Zener-St\"uckelberg transition\cite{STLandau-1932,STZener-1932,STStuckelberg-1932}.
The asymptotic behavior of the transition probability at $t \to +\infty$ is given by
\begin{eqnarray}
 P_{\rm LZS} = \exp\left( -\frac{(\Delta E)^2}{4v}\right),
\end{eqnarray}
where $\Delta E$ is the energy gap at $h=0$; in this case $\Delta E = 2\Gamma$.
The Landau-Zener-St\"uckelberg transition is adopted not only for single spin problem but also many-body quantum systems.
Actually, a couple of quantum dynamical behaviors can be analyzed by the Landau-Zener-St\"uckelberg transition since in many cases, the energy level structure can be approximated by that of single spin system.
Then, in order to consider nonadiabatic transition in the quantum annealing, the knowledge of the Landau-Zener-St\"uckelberg transition is useful.

\section{Quantum Field Response of Frustrated Ising Systems}

\begin{figure}[b]
 \begin{center}
  \psfig{file=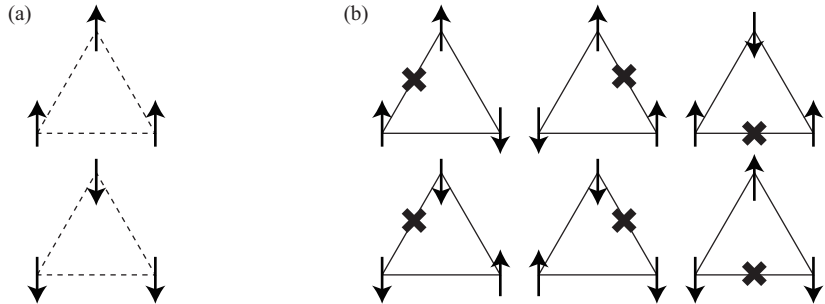,scale=1.}
 \end{center}
 \caption{
 Ground state spin configuration of the Ising model on a triangle cluster.
 The dotted and solid lines represent ferromagnetic and antiferromagnetic interactions, respectively.
 (a) Ferromagnetic case. All interactions are energetically favorable states.
 (b) Antiferromagnetic case. The crosses denote energetically unfavorable states.
 }
 \label{STfig:trianglecluster}
\end{figure}

In this section we consider transverse field response of frustrated Ising spin systems.
In order to explain the concept of frustration, we first consider the Ising model on a triangle cluster.
The Hamiltonian of this model is given by 
\begin{eqnarray}
 {\cal H}_{\vartriangle} = -J \left( \sigma_1^z \sigma_2^z + \sigma_2^z \sigma_3^z + \sigma_3^z \sigma_1^z \right),
  \qquad
  \sigma_i^z = \pm 1.
\end{eqnarray}
When the coupling constant $J$ is positive/negative, this is called ferromagnetic/antiferromagnetic interaction.
Figure \ref{STfig:trianglecluster} (a) shows the ground states in the case of ferromagnetic interaction.
In the ground states, all interactions are energetically favorable.
On the other hand, when the interactions are antiferromagnetic ($J<0$), in the ground state, there is a bond which is energetically unfavorable shown in Fig. \ref{STfig:trianglecluster}(b).
Such a nature is called frustration.
Since frustration prevents the system from ordering and makes peculiar density of states, unconventional order and characteristic dynamic behaviors appear\cite{STToulouse-1977,STLiebmann-1986,STKawamura-1998,STDiep-2005,STTanaka-2005,STTanaka-2007b,STTanaka-2007c,STTamura-2008,STTanaka-2009b,STTanaka-2010,STTamura-2011,STTamura-2011b}.

Frustration appears in antiferromagnet on triangle-based lattices such as triangular lattice, kagom\'e lattice, and pyrochlore lattice.
In random Ising spin systems, frustration also appears randomly on the lattice.
Since many optimization problems can be mapped onto random Ising models as mentioned above, it is an important issue to investigate quantum field response of frustrated systems comparing with thermal fluctuation effect.

In this paper we focus on two examples of frustrated Ising spin systems.
In Sec.~\ref{STsec:orderbydisorder}, we review transverse field effect on fully frustrated systems.
In Sec.~\ref{STsec:reentrant}, we study decorated bond systems in which correlation function behaves nonmonotonic against temperature and transverse magnetic field.

\subsection{Order by Disorder Effect in Fully Frustrated Systems}
\label{STsec:orderbydisorder}

\begin{figure}[t]
 \begin{center}
  \psfig{file=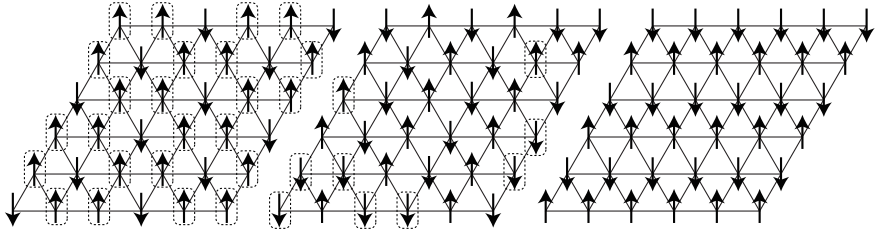,scale=0.8}
 \end{center}
 \caption{
 Typical ground states of antiferromagnetic Ising spin system on triangular lattice.
 The dotted boxes represent free spin where the internal field from the nearest neighbor sites are zero.
 }
 \label{STfig:GStrikagome}
\end{figure}

Here we consider fully frustrated systems where all plaquettes are frustrated.
There are macroscopically degenerated ground states in the fully frustrated systems\cite{STHusimi-1949,STSyozi-1949,STHoutappel-1950,STWannier-1950,STKano-1953,STWannier-1973}.
Typical configurations of ground state of the antiferromagnetic system on the triangular lattice are shown in Fig.~\ref{STfig:GStrikagome}.
In Fig.~\ref{STfig:GStrikagome}, we adopt the periodic boundary condition.
The dotted boxes in these figures indicate ``free spin'' where the internal field from the nearest neighbor sites is zero.
The number of free spins is a useful way to represent a character of each configuration.
From left to right in Fig.~\ref{STfig:GStrikagome}, the number of free spins decreases.
Here we first consider simulated annealing for such fully frustrated systems.
From the principle of equal weight, all distinct degenerated ground states can be obtained with the same probability at $T=0+$ in the simulated annealing.

\begin{figure}[b]
 \begin{center}
  \psfig{file=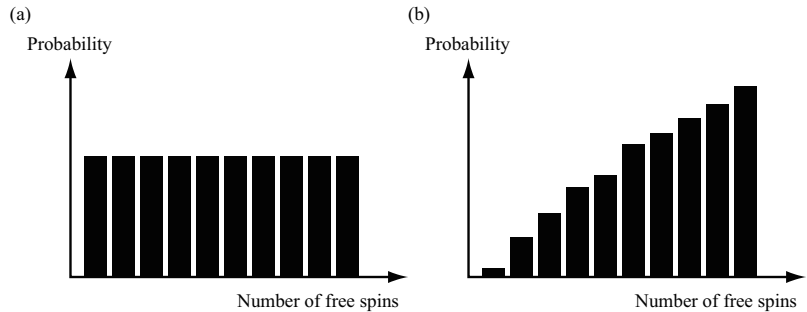,scale=0.9}
 \end{center}
 \caption{
 Schematic picture of probability distributions of the ground state for (a) simulated annealing and (b) quantum annealing.
 The number of free spins represents the configuration of ground states.
 }
 \label{STfig:schematicdis}
\end{figure}

By the way when we adopt quantum annealing for fully frustrated systems, whether does the probability distribution of the obtained ground states become flat or not?
If the probability distribution becomes biased distribution, which are states selected?
From the preceding studies, it is well-known that the states which have many free spins tend to appear in the quantum annealing\cite{STMatsuda-2009,STTanaka-2010book,STTanaka-2011a,STTanaka-2011c}.
The left panel in Fig.~\ref{STfig:GStrikagome} represents the maximum free spin state.
Actually the probability of this state is the maximum in the adiabatic limit.
This is because the transverse field is represented as
\begin{eqnarray}
 - \Gamma \sum_i \hat{\sigma}_i^x = -\Gamma \sum_i \left( \hat{\sigma}_i^+ + \hat{\sigma}_i^- \right),
\end{eqnarray}
and corresponds to sum of the spin-flip operator.
Then the states which have large number of free spins appear with high probability.
Schematic picture of the probability distributions of the ground state obtained by the simulated annealing and the quantum annealing is shown in Fig.~\ref{STfig:schematicdis}.
The maximum free spin states are ``ordered states'' which are mediated by quantum fluctuation effect.
Then, the nature is called ``order by disorder'' which is a famous feature in fully frustrated systems\cite{STLiebmann-1986,STKawamura-1998,STDiep-2005}.
The energy of the states obtained by simulated annealing with slow schedule is the same as that of the states obtained by quantum annealing.
Although the maximum free spin states are selected in this case, the other type of ground states can be selected when we adopt the other type of quantum fluctuation.
It is an interesting problem to investigate such probability distribution when we decrease temperature and transverse field with finite speed\cite{STTanaka-Tamurainprep}.

\subsection{Decorated Bond Systems}
\label{STsec:reentrant}

\begin{figure}[b]
 \begin{center}
  \psfig{file=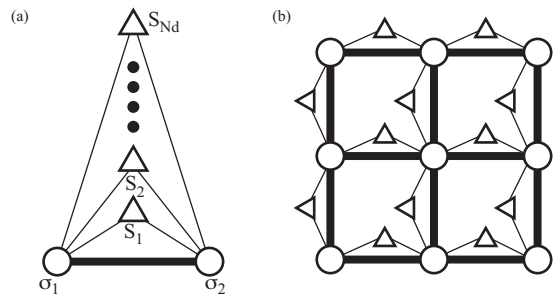,scale=1.}
 \end{center}
 \caption{
 (a) Unit of the decorated bond system. (b) Square lattice with decorated bond for $N_{\rm d}=1$.
 }
 \label{STfig:decstructure}
\end{figure}

Next we consider quantum field response of decorated bond systems.
Figure \ref{STfig:decstructure} shows a structure of decorated bond systems.
The circles and triangles in Fig.~\ref{STfig:decstructure} denote system spins and decorated spins, respectively.
Before we consider thermodynamic properties of lattice system with decorated bond shown in Fig.~\ref{STfig:decstructure} (b), we first consider temperature dependent correlation function between system spins of the decorated bond unit shown in Fig.~\ref{STfig:decstructure} (a).
The Hamiltonian of the unit is given as
\begin{eqnarray}
 {\cal H}_{\rm unit} = -J_{\rm dir} \sigma_1^z \sigma_2^z
  -J \sum_{i=1}^{N_{\rm d}} \left( \sigma_1^z + \sigma_2^z \right) s_i^z,
  \qquad
  \sigma_i^z = \pm 1, s_i^z = \pm 1,
\end{eqnarray}
where $J_{\rm dir}$ and $J$ represent direct coupling between system spins and decorated bond between a system spin and a decorated spin, respectively.

\begin{figure}[b]
 \begin{center}
  \psfig{file=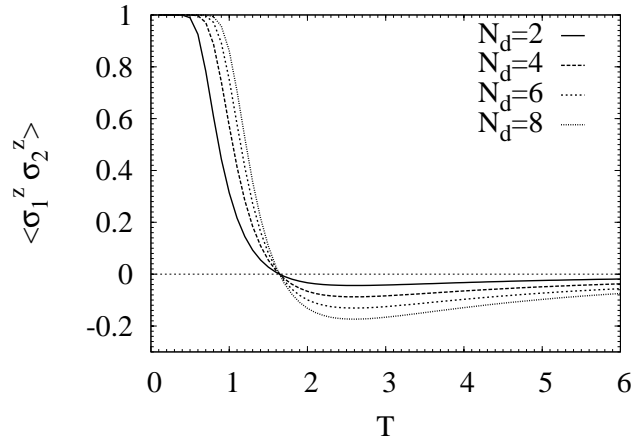,scale=0.7}
 \end{center}
 \caption{
 Correlation function between system spins as a function of temperature.
 }
 \label{STfig:cor_vs_T_nonmonotonic}
\end{figure}

The correlation function between system spins can be calculated exactly as
\begin{eqnarray}
 &&\left\langle \sigma_1^z \sigma_2^z \right\rangle = \tanh K_{\rm eff},\\
 &&{\rm Tr}_{\{s_i\}} {\rm e}^{-\beta {\cal H}} = A {\rm e}^{K_{\rm eff} \sigma_1^z \sigma_2^z},\\
 &&K_{\rm eff} = \frac{N_{\rm d}}{2} \log \cosh (2\beta J) + \beta J_{\rm dir},
\end{eqnarray}
where $A$ is an irrelevant factor.
Here if $J_{\rm dir}=0$ and $J>0$, the correlation function is always positive and monotonic decreasing function against temperature.
On the other hand, when $J=0$ and $J_{\rm dir}<0$, the correlation function is always negative and monotonic increasing function against temperature.
Then, by tuning the ratio $J_{\rm dir}$ and $J$ with keeping $J_{\rm dir}<0$ and $J>0$, the correlation function $\langle \sigma_1^z \sigma_2^z \rangle$ behaves non-monotonic as a function of temperature\cite{STFradkin-1976,STMiyashita-1983,STKitatani-1985,STKitatani-1986,STMiyashita-2001,STTanaka-2005,STTanaka-2010}.
In this paper we adopt $J_{\rm dir}=-\frac{N_{\rm d}}{2}J$ and $J$ as an energy unit.
The temperature dependency of correlation function in this case is shown in Fig.~\ref{STfig:cor_vs_T_nonmonotonic}.
It is noted that the signs of correlation function and the effective coupling $K_{\rm eff}$ have both plus and minus values depending on temperature.

In order to investigate the reason for such a non-monotonic behavior, we consider probability distributions of ferromagnetically correlated states ($\sigma_1^z \sigma_2^z = +1$) and antiferromagnetically correlated states ($\sigma_1^z \sigma_2^z = -1$).
The number of ferromagnetically correlated states and that of antiferromagnetically correlated states are $2^{N_{\rm d}+1}$.
Let $N_{\rm s}$ be the number of decorated spins in which directions of these spins are the same as $\sigma_1^z$.
The probability distributions of ferromagnetically correlated states $P_{\rm F}(N_{\rm s})$ and antiferromagnetically correlated states $P_{\rm AF}(N_{\rm s})$ are given by
\begin{eqnarray}
 &&P_{\rm F}(N_{\rm s}) =
  \frac{{\rm e}^{-\beta \frac{N_{\rm d}}{2}J}}{2\cosh \left( \beta\frac{N_{\rm d}}{2} J \right)}
  \left(
   \begin{array}{c}
    N_{\rm d}\\
    N_{\rm s}\\
   \end{array}
  \right)
  \left(
   \frac{1}{2}
  \right)^{N_{\rm d}}
  \frac{{\rm e}^{-2\beta(N_{\rm d}-2N_{\rm s})J}}{\cosh^{N_{\rm d}} \left( 2\beta J\right)},\\
 &&P_{\rm AF}(N_{\rm s}) =
  \frac{{\rm e}^{\beta \frac{N_{\rm d}}{2}J}}{2\cosh \left( \beta\frac{N_{\rm d}}{2} J \right)}
  \left(
   \begin{array}{c}
    N_{\rm d}\\
    N_{\rm s}\\
   \end{array}
  \right)
  \left(
   \frac{1}{2}
  \right)^{N_{\rm d}}.
\end{eqnarray}
It should be noted that the following relation is obviously satisfied:
\begin{eqnarray}
 \sum_{N_{\rm s}=0}^{N_{\rm d}} \left[ P_{\rm F}(N_{\rm s}) + P_{\rm AF}(N_{\rm s}) \right]=1.
\end{eqnarray}

At zero temperature all system spins and decorated spins are the same value.
Then the probability distribution is as follows: $P_{\rm F}(N_{\rm d})=1$ and otherwise are zero.
At the temperature where $\sum_{N_{\rm s}}P_{\rm F}(N_{\rm s}) = \sum_{N_{\rm s}}P_{\rm AF}(N_{\rm s})$ is satisfied, the correlation function $\langle \sigma_1^z \sigma_2^z \rangle$ becomes zero.
This is the reason why the correlation function and effective coupling behave non-monotonic.

Suppose we consider square lattice system with decorated bond for large enough $N_{\rm d}$.
In this case there is temperature region where the absolute value of effective coupling exceeds the critical value of the Ising spin system on square lattice $K_{\rm c}=\frac{1}{2}\log\left( 1 + \sqrt{2}\right)$\cite{STOnsager-1944}.
Then, paramagnetic phase $\to$ antiferromagnetic phase $\to$ paramagnetic phase $\to$ ferromagnetic phase appear as temperature decreases and at each phase boundary, phase transition takes place.
Such successive phase transitions are called reentrant phase transition.
Reentrant phase transitions and non-monotonic behavior of correlation function are typical nature of frustrated systems\cite{STAzaria-1987}.

\begin{figure}[t]
 \begin{center}
  \psfig{file=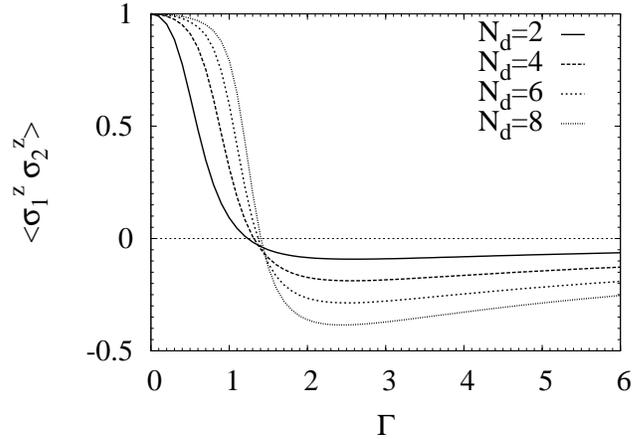,scale=0.7}
 \end{center}
 \caption{
 Correlation function between the system spins as a function of transverse field.
 }
 \label{STfig:cor_vs_Gamma_nonmonotonic}
\end{figure}

So far we showed thermal fluctuation response of decorated Ising spin systems.
Next we consider transverse field response of this system at zero temperature.
Then the Hamiltonian is given by
\begin{eqnarray}
 {\cal H} = -J_{\rm dir} \hat{\sigma}_1^z \hat{\sigma}_2^z 
  -J \sum_{i=1}^{N_{\rm d}} \left( \hat{\sigma}_1^z + \hat{\sigma}_2^z \right) \hat{s}_i^z
  -\Gamma \left( \hat{s}_1^x + \hat{s}_2^x + \sum_{i=1}^{N_{\rm d}} \hat{\sigma}_i^x \right),
\end{eqnarray}
where $\hat{s}_i^\alpha$ denotes the $\alpha$-element of the Pauli matrix of the $i$-th decorated spin.
As the previous example, we adopt $J_{\rm dir}=-\frac{N_{\rm d}}{2}J$.
We calculate the correlation function between system spins along the $z$-axis at zero temperature as a function of transverse field as shown in Fig.~\ref{STfig:cor_vs_Gamma_nonmonotonic}.
As well as the thermal fluctuation, the correlation function behaves non-monotonic as a function of transverse field.
This result indicates there is a similar point between thermal fluctuation and quantum fluctuation originated by transverse field.
Suppose we consider the system where reentrant phase transition occurs when we decrease temperature.
If we adopt the simulated annealing method, we face on the difficulty which comes from the critical slowing down and the simulated annealing is not useful.
Whenever we adopt the quantum annealing using time-dependent transverse field, the system also exhibits reentrant phase transition.
Then we should consider other type of quantum fluctuation effect in order to avoid the occurrence of the phase transition.

\section{Conclusion and Future Perspective}

In this paper, we reviewed how to implement quantum annealing focusing on the methods based on quantum Monte Carlo method and time-development Schr\"odinger equation.
The quantum annealing method is expected to be a powerful tool to obtain the best solution of optimization problems.
This method is a general method and can be implemented easily.
In this method we use quantum fluctuation to find the stable state.
In order to put into practical use, it is important to investigate quantum fluctuation effects for optimization problems.
Then we considered quantum fluctuation effect for frustrated systems by taking the transverse Ising model as an example, since the Ising model which represents optimization problem is a frustrated system in general.
We explained differences and similarities between thermal fluctuation and quantum fluctuation.
By building up the organized knowledge of the thermal fluctuation and the quantum fluctuation, the quantum annealing method develops into a truly useful algorithm.

\section*{Acknowledgement}

The authors are grateful to Bernard Barbara, Masaki Hirano, Yoshiki Matsuda, Seiji Miyashita, Hans de Raedt, Per Arne Rikvold, and Eric Vincent for their valuable comments.
R.T. is partly supported by Global COE Program ``the Physical Sciences Frontier'', MEXT, Japan.
S.T. is partly supported by Grand-in-Aid for JSPS Fellows (23-7601).
The computation in the present work was performed on computers at the Supercomputer Center, Institute for Solid State Physics, University of Tokyo. 

\bibliographystyle{ws-procs9x6}
\bibliography{ws-pro-sample}

\begin{thebibliography}{999}
\bibitem{STIsing-1925}
	E.~Ising, 
	{\em Z. Physik} {\bf 31}, 253 (1925).

\bibitem{STMezard-1987}
	M.~M\'ezard, G.~Parisi, and M.~A.~Virasoro,
	{\it Spin Glass Theory and Beyond} (World Scientific, 1987).

\bibitem{STFischer-1993}
	K.~H.~Fischer and J.~A.~Hertz,
	{\it Spin Glasses} (Cambridge University Press, 1993).

\bibitem{STYoung-1998}
	A.~P.~Young,
	{\it Spin Glasses and Random Fields} (World Scientific, 1998).

\bibitem{STKirkpatrick-1983}
	S.~Kirkpatrick, C.~D.~Gelatt Jr., and M.~P.~Vecchi, 
	{\em Science} {\bf 220}, 671 (1983).

\bibitem{STKirkpatrick-1984}
	S.~Kirkpatrick, 
	{\em J. Stat. Phys.} {\bf 34}, 975 (1984).

\bibitem{STGeman-1984}
	S.~Geman and D.~Geman,
	{\em IEEE Transactions on Pattern Analysis and Machine Intelligence} {\bf 6}, 721 (1984).

\bibitem{STFinnila-1994}
	A.~B.~Finnila, M.~A.~Gomez, C.~Sebenik, C.~Stenson, and J.~D.~Doll,
	{\em Chem. Phys. Lett.} {\bf 219} 343 (1994).

\bibitem{STKadowaki-1998}
	T.~Kadowaki and H.~Nishimori,
	{\em Phys. Rev. E} {\bf 58}, 5355 (1998).

\bibitem{STBrooke-1999}
	J.~Brooke, D.~Bitko, T.~F.~Rosenbaum, and G.~Aeppli,
	{\em Science} {\bf 284}, 779 (1999).

\bibitem{STFarhi-2001}
	E.~Farhi, J.~Goldstone, S.~Gutmann, J.~Lapan, A.~Lundgren, and D.~Preda,
	{\em Science} {\bf 292}, 472 (2001).

\bibitem{STSantoro-2002}
	G.~E.~Santoro, R.~Marto\v{n}\'ak, E.~Tosatti, and R.~Car,
	{\em Science} {\bf 295}, 2427 (2002).

\bibitem{STMatronak-2004}
	R.~Marton\'{a}k, G.~E.~Santoro, and E.~Tosatti,
	{\em Phys. Rev. E} {\bf 70}, 057701 (2004).

\bibitem{STBattaglia-2005}
	D.~A.~Battaglia, G.~E.~Santoro, and E.~Tosatti,
	{\em Phys. Rev. E} {\bf 71}, 066707 (2005).

\bibitem{STSuzuki-2005}
	S.~Suzuki and M.~Okada,
	{\em J. Phys. Soc. Jpn.} {\bf 74}, 1649 (2005).

\bibitem{STDas-2005}
	A.~Das and B.~K.~Chakrabarti,
	{\it Quantum Annealing and Related Optimization Methods} (Springer, 2005).

\bibitem{STTanaka-2007}
	S.~Tanaka and S.~Miyashita,
	{\em J. Magn. Magn. Mater.} {\bf 310}, e468 (2007).

\bibitem{STDas-2008}
	A.~Das and B.~K.~Chakrabarti,
	{\em Rev. Mod. Phys.} {\bf 80}, 1061 (2008).

\bibitem{STKurihara-2009}
	K.~Kurihara, S.~Tanaka, and S.~Miyashita,
	{\em Proceedings of the 25th Conference on Uncertainty in Artificial Intelligence} (2009).

\bibitem{STSato-2009}
	I.~Sato, K.~Kurihara, S.~Tanaka, H.~Nakagawa, and S.~Miyashita,
	{\em Proceedings of the 25th Conference on Uncertainty in Artificial Intelligence} (2009).
	
\bibitem{STMorita-2009}
	S.~Morita, S.~Suzuki, and T.~Nakamura,
	{\it Phys. Rev. E} {\bf 79}, 065701(R) (2009).

\bibitem{STInoue-2010}
	J.~Inoue, Y.~Saika, and M.~Okada,
	{\it Lecture Note in Physics ``Quantum Quenching, Annealing, and Computation''} (Springer) {\bf 802}, 283 (2010).

\bibitem{STTanaka-2009}
	S.~Miyashita, S.~Tanaka, H.~de Raedt, and B.~Barbara,
	{\em J. Phys.: Conf. Ser.} {\bf 143}, 012005 (2009).

\bibitem{STTanaka-2010}
	S.~Tanaka and S.~Miyashita,
	{\em Phys. Rev. E} {\bf 81}, 051138 (2010).

\bibitem{STTanaka-2010book}
	S.~Tanaka, M.~Hirano, and S.~Miyashita,
	{\it Lecture Note in Physics ``Quantum Quenching, Annealing, and Computation''} (Springer) {\bf 802}, 215 (2010).

\bibitem{STChandra-2010}
	A.~K.~Chandra, A.~Das, J.~Inoue, and B.~K.~Chakrabarti,
	{\it Lecture Note in Physics ``Quantum Quenching, Annealing, and Computation''} (Springer) {\bf 802}, 235 (2010).

\bibitem{STOhzeki-2011}
	M.~Ohzeki and H.~Nishimori,
	{\em J. Comp. and Theor. Nanoscience} {\bf 8}, 963 (2011).

\bibitem{STTanaka-2011a}
	S.~Tanaka, M.~Hirano, and S.~Miyashita,
	{\em Physica E} {\bf 43}, 766 (2010).

\bibitem{STTanaka-2011b}
	S.~Tanaka, R.~Tamura, I.~Sato, and K.~Kurihara,
	to appear in {\it Kinki University Quantum Computing Series: ``Summer School on Diversities in Quantum Computation/Information''}.

\bibitem{STTanaka-2011c}
	S.~Tanaka,
	to appear in {\it proceedings of Kinki University Quantum Computing Series: ``Symposium on Quantum Information and Quantum Computing''} (2011).

\bibitem{STSwendsen-1987}
	R.~H.~Swendsen and J.~S.~Wang,
	{\em Phys. Rev. Lett.} {\bf 58}, 86 (1987).

\bibitem{STWolff-1989}
	U.~Wolff,
	{\em Phys. Rev. Lett.} {\bf 62}, 361 (1989).

\bibitem{STHukushima-1996}
	K.~Hukushima and K.~Nemoto,
	{\em J. Phys. Soc. Jpn.} {\bf 65}, 1604 (1996).

\bibitem{STHarada-2004}
	N.~Kawashima and K.~Harada,
	{\em J. Phys. Soc. Jpn.} {\bf 73}, 1379 (2004).

\bibitem{STNakamura-2008}
	T.~Nakamura,
	{\em Phys. Rev. Lett.} {\bf 101}, 210602 (2008).

\bibitem{STSuwa-2010}
	H.~Suwa and S.~Todo,
	{\em Phys. Rev. Lett.} {\bf 105}, 120603 (2010).

\bibitem{STSuwa-2011}
	H.~Suwa and S.~Todo,
	{\em arXiv}:1106.3562.

\bibitem{STSuzuki-2007}
	S.~Suzuki and M.~Okada,
	{\em Interdisciplinary Information Sciences} {\bf 13}, 49 (2007).

\bibitem{STTanaka-2000}
	K.~Tanaka and T.~Horiguchi,
	{\em Electronics and Communications in Japan} {\bf 83}, 84 (2000).

\bibitem{STBloch-2008}
	I.~Bloch, J.~Dalibard, W.~Zwerger,
	{\em Rev. Mod. Phys.} {\bf 80}, 885 (2008).

\bibitem{STKim-2010}
	K.~Kim, M.~-S.~Chang, S.~Korenblit, R.~Islam, E.~E.~Edwards, J.~K.~Freericks, G.~-D.~Lin, L.~-M.~Duan, and C.~Monroe,
	{\em Nature} {\bf 465}, 590 (2010).

\bibitem{STMa-2011}
	X.~-S.~Ma, B.~Dakic, W.~Naylor, A.~Zeilinger, and P.~Walther,
	{\em Nat. Phys.} {\bf 7}, 399 (2011).

\bibitem{STSimon-2011}
	J.~Simon, W.~S.~Bakr, R.~Ma, M.~E.~Tai, P.~M.~Preiss, and M.~Greiner,
	{\em Nature} {\bf 472}, 307 (2011).

\bibitem{STStruck-2011}
	J.~Struck, C.~\"Olschl\"ager, R.~L.~Targat, P.~S.~Panahi, A.~Eckardt, M.~Lewenstein, P.~Windpassinger, and K.~Sengstock,
	{\em Science} {\bf 333}, 996 (2011).

\bibitem{STTrotter-1959}
	H.~F.~Trotter,
	{\em Proceedings of the American Mathematical Society} {\bf 10}, 545 (1959).

\bibitem{STSuzuki-1976}
	M.~Suzuki,
	{\em Prog. Theor. Phys.} {\bf 56}, 1454 (1976).

\bibitem{STLandau-1932}
	L.~Landau,
	{\em Phys. Z. Souwjetunion} {\bf 2}, 46 (1932).

\bibitem{STZener-1932}
	C.~Zener,
	{\em Proc. R. Soc. London Ser. A} {\bf 137}, 696 (1932).

\bibitem{STStuckelberg-1932}
	E.~C.~G.~St\"uckelberg,
	{\em Helv. Phys. Acta} {\bf 5}, 369 (1932).

\bibitem{STToulouse-1977}
	G.~Toulouse,
	{\em Commun. Phys.} (London) {\bf 2}, 115 (1977).

\bibitem{STLiebmann-1986}
	R.~Liebmann,
	{\it Statistical Mechanics of Periodic Frustrated Ising Systems} (Springer-Verlag, Berlin/Heidelberg, GmbH, Heidelberg, 1986).

\bibitem{STKawamura-1998}
	H.~Kawamura,
	{\em J. Phys.: Condens. Matter} {\bf 10}, 4707 (1998).

\bibitem{STDiep-2005}
	H.~T.~Diep (ed.),
	{\it Frustrated Spin Systems} (World Scientific, Singapore, 2005).

\bibitem{STTanaka-2005}
	S.~Tanaka and S.~Miyashita,
	{\em Prog. Theor. Phys. Suppl.} {\bf 157}, 34 (2005).

\bibitem{STTanaka-2007b}
	S.~Miyashita, S.~Tanaka, and M.~Hirano,
	{\em J. Phys. Soc. Jpn.} {\bf 76}, 083001 (2007).

\bibitem{STTanaka-2007c}
	S.~Tanaka and S.~Miyashita,
	{\em J. Phys. Soc. Jpn.} {\bf 76}, 103001 (2007).

\bibitem{STTamura-2008}
	R.~Tamura and N.~Kawashima,
	{\em J. Phys. Soc. Jpn.} {\bf 77}, 103002 (2008).

\bibitem{STTanaka-2009b}
	S.~Tanaka and S.~Miyashita,
	{\em J. Phys. Soc. Jpn.} {\bf 78}, 084002 (2009).

\bibitem{STTamura-2011}
	R.~Tamura and N.~Kawashima,
	{\em J. Phys. Soc. Jpn.} {\bf 80}, 074008 (2011).

\bibitem{STTamura-2011b}
	R.~Tamura, N.~Kawashima, T.~Yamamoto, C.~Tassel, and H.~Kageyama,
	{\em Phys. Rev. B} {\bf 84}, 214408 (2011).

\bibitem{STMatsuda-2009}
	Y.~Matsuda, H.~Nishimori, and H.~G.~Katzgraber,
	{\em New J. Phys.} {\bf 11}, 073021 (2009).

\bibitem{STHusimi-1949}
	K.~Husimi and I.~Syozi,
	{\em Prog. Theor. Phys.} {\bf 5}, 177 (1949).

\bibitem{STSyozi-1949}
	I.~Syozi,
	{\em Prog. Theor. Phys.} {\bf 5}, 341 (1949).

\bibitem{STHoutappel-1950}
	R.~M.~F.~Houtappel,
	{\em Physica} {\bf 16}, 425 (1950).

\bibitem{STWannier-1950}
	G.~H.~Wannier,
	{\em Phys. Rev.} {\bf 79}, 357 (1950).

\bibitem{STKano-1953}
	K.~Kano and S.~Naya,
	{\em Prog. Theor. Phys.} {\bf 10}, 158 (1953).

\bibitem{STWannier-1973}
	G.~H.~Wannier,
	{\em Phys. Rev. B} {\bf 7}, 5017 (1973).

\bibitem{STTanaka-Tamurainprep}
	S.~Tanaka and R.~Tamura,
	{\em in preparation.}

\bibitem{STFradkin-1976}
	E.~H.~Fradkin and T.~P.~Eggarter, 
	{\em Phys. Rev. A} {\bf 14}, 495 (1976).

\bibitem{STMiyashita-1983}
	S.~Miyashita,
	{\em Prog. Theor. Phys.} {\bf 69}, 714 (1983).

\bibitem{STKitatani-1985}
	H.~Kitatani, S.~Miyashita, and M.~Suzuki,
	{\em Phys. Lett.} {\bf 108A}, 45 (1985).

\bibitem{STKitatani-1986}
	H.~Kitatani, S.~Miyashita, and M.~Suzuki,
	{\em J. Phys. Soc. Jpn.} {\bf 55}, 865 (1986).

\bibitem{STMiyashita-2001}
	S.~Miyashita and E.~Vincent,
	{\em Eur. Phys. J. B} {\bf 22}, 203 (2001).

\bibitem{STOnsager-1944}
	L.~Onsager,
	{\em Phys. Rev.} {\bf 65}, 117 (1944).

\bibitem{STAzaria-1987}
	P.~Azaria, H.~T.~Diep, and H.~Giacomini,
	{\em Phys. Rev. Lett.} {\bf 59}, 1629 (1987).

\end{thebibliography}

\end{document}